\documentclass[preprint,12pt]{elsarticle}

\usepackage{graphicx,epsfig}
\usepackage{graphicx,epsfig}
\usepackage{amsmath}
\usepackage{amsfonts}
\usepackage{amssymb}
\usepackage{subfigure}
\usepackage{color}
\usepackage{ulem} 
\graphicspath{{./}{../}{./figures/}}

\def\fun#1#2{\lower3.6pt\vbox{\baselineskip0pt\lineskip.9pt
  \ialign{$\mathsurround=0pt#1\hfil##\hfil$\crcr#2\crcr\sim\crcr}}}

\newcommand{\x}{\textbf{x}}

\newcommand{\Id}{\textbf{I}}
\newcommand{\F}{\mathbf{F}}
\newcommand{\U}{\mathbf{U}}

\newcommand{\f}{\mathbf{f}}

\newcommand{\tr}{{\rm tr}\;}

\renewcommand{\d}{\partial}

\renewcommand{\u}{\textbf{u}}

\newcommand\X{\mathbf{X}}
\newcommand{\C}{\mathbf{C}}

\newcommand{\tauB}{\boldsymbol{\tau}}
\newcommand{\srt}{\dot{\boldsymbol{\gamma}}}
\newcommand{\Str}{\mathcal{S}}
\renewcommand{\v}{\mathbf{v}}

\newcommand{\sig}{\boldsymbol{\sigma}}

\newcommand{\gdot}{\dot{\gamma}}

\newcommand{\De}{\textrm{De}}     
\newcommand{\Wi}{\textrm{Wi}}     

\definecolor{review_color}{rgb}{0.0, 0.0, 0.0}  
\newcommand{\change}[1]{\textcolor{review_color}{#1}}

\begin{document}


%




\begin{frontmatter}

\title{Polymer stress growth in viscoelastic fluids in oscillating
  extensional flows with applications to micro-organism locomotion}



\author[bt]{Becca Thomases\corref{co}}
\cortext[co]{Corresponding author}
\ead{thomases@math.ucdavis.edu}

\author[bt]{Robert D. Guy}
\ead{guy@math.ucdavis.edu}

\address[bt]{Department of Mathematics, University of California, Davis, CA 95616}


\date{\today}

\begin{abstract}
Simulations of undulatory swimming in viscoelastic fluids with large amplitude gaits show
concentration of polymer elastic stress at the tips of the swimmers.
%
%
We use a series of related theoretical investigations to probe the origin of these  concentrated stresses.
First the polymer
stress is computed analytically at a given oscillating extensional
stagnation point in a viscoelastic fluid. 
The
analysis identifies a Deborah number ($\De$) dependent Weissenberg number ($\Wi$)
transition below which the stress is linear in $\Wi,$ and above which
the stress grows exponentially in $\Wi.$
Next, 
stress and velocity are found from numerical simulations in an 
oscillating 4-roll mill geometry. 
%
%
%
 %
  %
 The stress from these simulations is
compared with the theoretical calculation of stress in the decoupled (given flow) case, and similar stress behavior is observed.
The flow around tips of a time-reversible flexing filament in a
viscoelastic fluid is shown to exhibit an oscillating extension along
particle trajectories, and the stress response exhibits similar
transitions.
However in the high amplitude, high $\De$ regime the stress feedback
on the flow leads to non time-reversible particle trajectories that
experience asymmetric stretching and compression, and the stress
grows more significantly in this regime. These results help explain past observations of large stress
concentration for large amplitude swimmers and non-monotonic
dependence on $\De$ of swimming speeds.

%

\end{abstract}


\end{frontmatter}


%
\section{Introduction}

Simulations of swimming in viscoelastic fluids involving large amplitude gaits
\cite{teran2010viscoelastic,thomases2017role,spagnolie2013locomotion} show substantially
different swimming speeds than those found in low amplitude simulations and asymptotic
analyses
\cite{fu2007theory,fu2008beating,LAUGA:2007,fu2009swimming,riley2014enhanced,elfring2016effect}.
%
Concentration of polymer elastic stress at the tips of slender objects
has been seen in numerical simulations of flagellated swimmers in
viscoelastic fluids
\cite{teran2010viscoelastic,thomases2014mechanisms,thomases2017role,li2017flagellar},
and it is thought that the presence of these large stresses is related to the observed
differences in behavior at low and high amplitude. 
We recently explained the origin of the stress concentration at the tips of steady, translating cylinders \cite{li2019orientation}.
The tips of swimmers, however, experience unsteady oscillating motion. 
This paper analyzes oscillatory extensional flows, which are similar
to the flows around bending objects, and it identifies parameter regimes which 
lead to the presence of large concentrated stresses.

\begin{figure}
  \centering
  \includegraphics[width=0.85\textwidth]{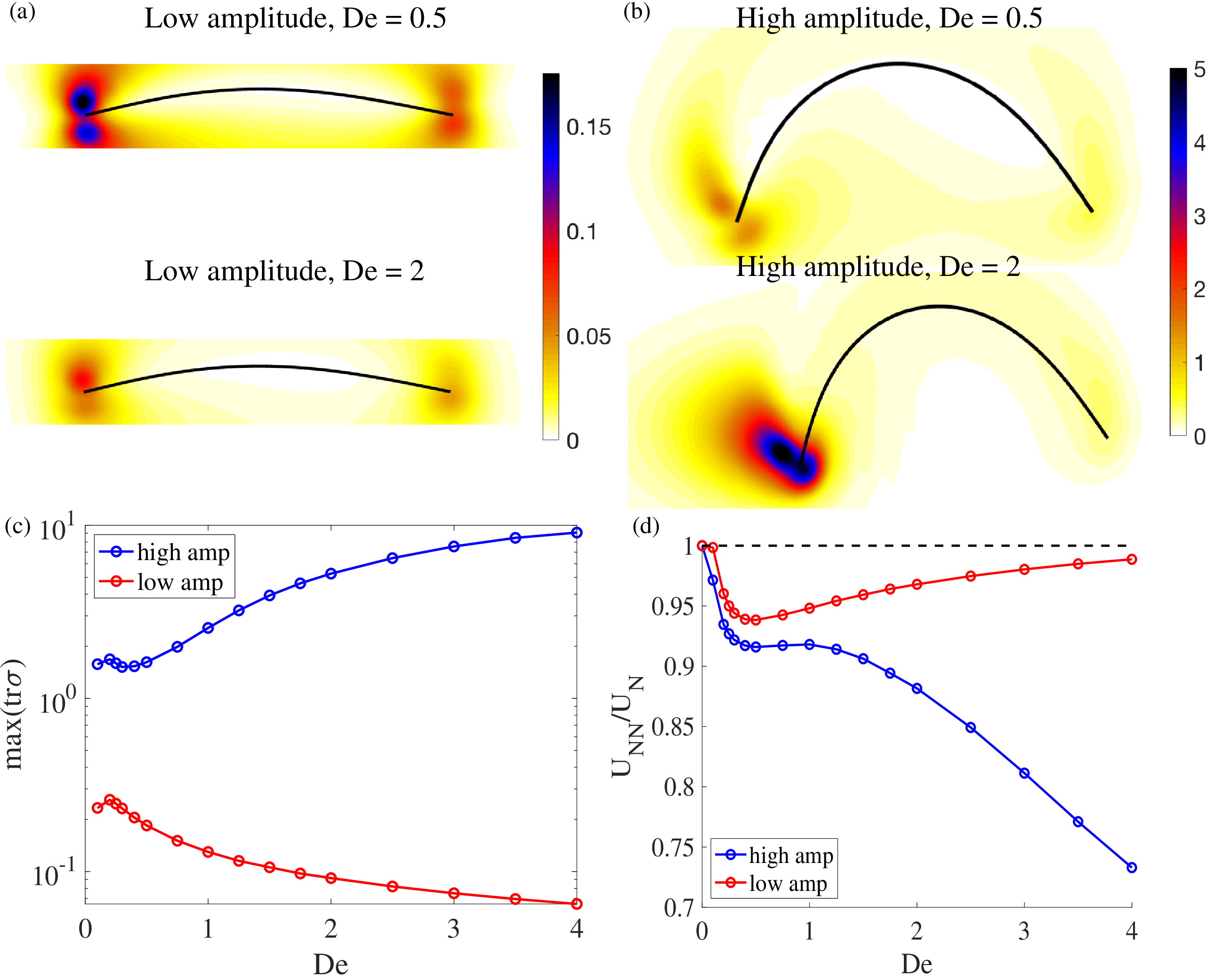}
  \caption{(a,b) Snapshots of the stress distributions around
    swimmers with (a) low and (b) high amplitude gaits at both low and
    high Deborah numbers ($\De$).  The swimmer gait is prescribed
   with a curvature $\kappa(s,t)=\left(A_t(1-s)+A_hs\right)\sin(2\pi t+\pi s),$
  where $(A_t,A_h)$ are the tail and head curvature amplitudes, and $(A_t,A_h)=(5,2)$
  for the large amplitude case and $(1,2/5)$ for low amplitude.
    The color field shows polymer strain
    energy density scaled by $2\mu_pA_t.$ A similar scaling will be used for the
    flexors in Sec.\ \ref{flexor:sec}. (c) Scaled strain energy
    density in a neighborhood near the tail as a function of $\De.$
    (d) Non-Newtonian swimming speed normalized by swimming speed in a
    Newtonian fluid.}
  \label{fig:TWALR}
\end{figure}

In Fig.\ \ref{fig:TWALR} we present results similar to those from
\cite{thomases2014mechanisms,thomases2017role} which compare low and
high amplitude undulatory swimmers in a 2D Stokes-Oldroyd-B fluid.  In
Fig.\ \ref{fig:TWALR} (a)-(b) we show the scaled strain energy density
(trace of the stress) for both low and high Deborah number and amplitude.
\change{The Deborah number is the
polymer relaxation time scaled by the flow time-scale, and is denoted by $\De.$}
Large stress accumulation at the tail only occurs in the high $\De,$
high amplitude case. In Fig.\ \ref{fig:TWALR}(c)-(d) we plot the
maximum strain energy density and swimming speed \change{(normalized by the swimming speed in a Newtonian fluid)} as a
function of $\De.$ In both the stress response and normalized swimming
speed, the high amplitude behavior is very different from the low
amplitude behavior at high $\De.$ At low $\De,$ low and high amplitude
motion results in similar normalized swimming speed, but significant
slow downs are seen for the high amplitude swimmers where the stress
is also very high.

Translating cylinders in a viscoelastic fluid
exhibit a Weissenberg number transition from low to high stress
concentration at the cylinder tips \cite{li2019orientation}.  The Weissenberg number, $\Wi,$ is the
polymer relaxation time scaled by the flow strain rate. This
transition is similar to the coil-stretch transition found for
viscoelastic fluids at steady extensional points.  Undulatory swimmers
are oscillating as well as translating, and the Deborah number is typically
reported as the relevant non-dimensional relaxation time in this case.
Here we show that the fluid near tips of oscillating filaments in 2D
experience oscillating extension along particle trajectories, and both
Deborah number and Weissenberg number are important. Our results extend known transitions in $\Wi$ at steady extensional stagnation
points to oscillating extensional points, and the $\Wi$ transition
becomes $\De$ dependent in this case.
The need to report both $\De$ and $\Wi$ has been well appreciated in the engineering community, see \cite{dealy2010weissenberg,poole2012deborah}
for nice discussions of these two parameters, but it has not been noted before in the context of micro-organism locomotion in viscoelastic fluids.

In this paper we analyze the stress response at a fixed oscillatory
extensional stagnation point with no stress feedback on the flow. We
compare these analytical results with different numerical simulations
in which the stress and flow are coupled.  We examine flow-stress
coupling in oscillating extension by forcing the flow with a 4-roll
mill background force that is oscillatory in time. Next, we look at
the flow around flexing filaments with a time-reversible oscillation
of a circular arc of a given amplitude. The flow around these
so-called \textit{flexors} is similar to the flow around undulatory
swimmers and provides a connection between the analysis of stress
response at oscillatory extensional stagnation points and recent
numerical studies on stress accumulation at tips of flagellated and
undulatory swimmers
\cite{teran2010viscoelastic,thomases2014mechanisms,salazar2016numerical,thomases2017flexible,li2017flagellar}.

\section{Model Equations}

We use the Oldroyd-B model of a viscoelastic fluid in the
zero-Reynolds number limit. The Oldroyd-B model is the simplest model
of a viscoelastic fluid which captures elastic effects such as storage
of memory from past deformation on a characteristic time-scale
$\lambda.$ We study zero-Reynolds number because this work is
motivated by micro-organism locomotion.
The model equations for velocity $\u,$ pressure $p,$ and polymer
stress tensor $\tauB$ are
\begin{gather}
  \label{stokes}-\nabla p+\mu_s\Delta\u+\nabla\cdot\tauB + \f = 0\\
  \label{ic}\nabla\cdot\u =0\\
  \label{ucderiv}\tauB+\lambda\stackrel{\nabla}\tauB =2\mu_p\srt,
\end{gather}
where $\stackrel{\nabla}\tauB,$ denotes the upper convected derivative, and is
defined by
\begin{equation}
  \label{ucderivdef}
  \stackrel{\nabla}\tauB=\d_t\tauB+\u\cdot\nabla\tauB-\left(\nabla\u\tauB+\tauB\nabla\u^T\right).
\end{equation} The solvent and polymer viscosities are  $\mu_s,$ and $\mu_p$, respectively, and
 $\srt=\left(\nabla\u+\nabla\u^T\right)/2$ is the rate of
strain tensor. The function $\f$ is the external forcing that drives the  system, and will be given explicitly for the different examples 
in Sec. \ref{sec_4roll} for the 4-roll mill geometry and in Sec. \ref{flexor:sec} for the flexor simulations. 


\section{Stress response to oscillating extension with no coupling}\label{sec_theory}

It is well known that in the Oldroyd-B model of a viscoelastic fluid,
the stress shows unbounded exponential growth at extensional points
when stretching outpaces relaxation. The rapid stretching originates
from the nonlinearity in the upper-convected derivative,
Eq.\ \eqref{ucderivdef}.  This derivative is the frame-invariant
material derivative of a tensor, or derivative of a tensor along a
particle path, making it essential in continuum models of viscoelastic
fluids.  Other similar models such as Giesekus, FENE-P, and PTT also
lead to large concentrated stresses in these regions of strong
extension \cite{guy2015computational}. We use the Oldroyd-B model here
because it is the simplest model that contains these nonlinear
features and is amenable to analysis. 

We begin by repeating the well known calculation for stress divergence
in the Oldroyd-B model for a fixed velocity, in order to frame the
following sections. Consider a linear extensional flow
$\u=\alpha(x,-y)$ for $\alpha>0$. At the origin, the diagonal
components of the stress satisfy
\begin{align}
  \label{tau11ode1} \lambda\frac{d}{dt}\tauB_{11} & =  \phantom{-}2\alpha\mu_p -(1-2\lambda\alpha)\tauB_{11} \\
  \label{tau22ode1} \lambda\frac{d}{dt}\tauB_{22}  & = -2\alpha\mu_p -(1+2\lambda\alpha)\tauB_{22}.
\end{align}
For $2\lambda\alpha <1$, the stress approaches a bounded steady
state, and for $2\lambda\alpha>1$, $\tauB_{11}$ grows exponentially in
time without bound. For this flow, the Weissenberg number is
$\Wi=2\alpha\lambda$, and the unbounded stress growth occurs for
$\Wi>1.$

We extended this classical result to the situation in which the strength
of extension is oscillating in time with mean zero and period $T$. Specifically, we
assume a velocity field (independent of the stress) of the form
\begin{equation}
  \label{oscext1}
  \u = \alpha h(t/T)(x,-y),
\end{equation}
where $h(t)$ is a periodic function with period $1$, mean zero, and
maximum 1. At the origin, the diagonal components of the stress
satisfy the following ODE's:
\begin{align} 
\label{tau11ode} \lambda\left(\frac{d}{dt}\tauB_{11}-2\alpha h(t/T)\tauB_{11}\right)+\tauB_{11}&=\phantom{-}2\alpha \mu_p h(t/T) \\ 
\label{tau22ode} \lambda\left(\frac{d}{dt}\tauB_{22}+2\alpha h(t/T)\tauB_{22}\right)+\tauB_{22}&=-2\alpha\mu_p h(t/T). 
\end{align}
We nondimensionalize these equations by scaling stress by $2\mu_p
\alpha$ and scaling time by the period $T$. We denote the
dimensionless stress by $\sig = \tauB/2\mu_p \alpha$. The dimensionless equations are
\begin{align} 
  \label{sig11ode} \De\;\frac{d}{dt}\sig_{11}+(1-h(t)\Wi)\;\sig_{11}&=\phantom{-}h(t) \\ 
  \label{sig22ode} \De\;\frac{d}{dt}\sig_{22}+(1+h(t)\Wi)\;\sig_{22}&=-h(t), 
\end{align}
where, as before, the Weissenberg number is $\Wi = 2\alpha\lambda,$ and
the Deborah number is $\De=\lambda/T$.

To gain insight from an analytic solution to these equations, we
choose the function $h$ to be the square wave
\begin{equation}
  h(t) = \begin{cases}
      \phantom{-}1 & \text{for }  \text{mod}(t,1) \leq 1/2 \\
     -1 &  \text{for } \text{mod}(t,1) > 1/2 
         \end{cases}.
\end{equation}
\change{This choice of $h$ permits us to find the periodic solution analytically, from which} we
 compute the
maximum in time of the trace of the stress (strain energy density) as
\begin{equation}\label{sig_theory}
   \max \tr\sig = 
   \frac{ 2\sinh\left(\frac{\Wi}{2\De}\right) 
         -2\Wi\sinh\left(\frac{1}{2\De}\right)}
        {(\Wi^2-1)\sinh\left(\frac{1}{2\De}\right)}.
\end{equation}
Unlike the case of steady extensional flow, the solution remains
bounded in time, and it approaches a periodic solution for all
$\Wi$. 

\begin{figure}
  \centering
  \includegraphics[width=0.95\textwidth]{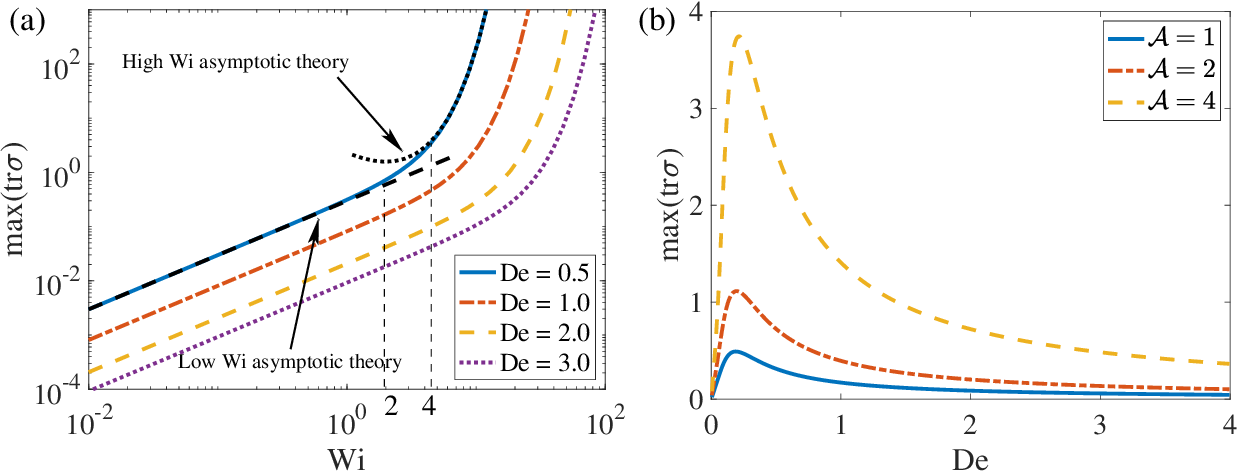}
  \caption{(a) Analytical solution of $\max(\tr\sig)$ as a function of $\Wi$ for $\De = 0.5,1,2,3$ at oscillatory extensional stagnation points with no stress feedback in the Stokes-Oldroyd-B model. Linear and exponential asymptotic approximations are included to highlight transition. $\Wi$ transition depends on $\De.$ (b) Analytical solution of $\max(\tr\sig)$ as a function of $\De$ for non-dimensional stretch rate $\mathcal{A}=1,2,4.$  }
  \label{fig:LAOE_sqwv}
\end{figure}

Figure \ref{fig:LAOE_sqwv}(a) shows how the stress depends on $\Wi$
for different fixed $\De$. This plot shows two different regimes for
how the stress depends on $\Wi$, and there is a Deborah number dependent
transition between the two regimes. To understand the behavior in the
two regimes, we expand the max trace of the stress in the limits of
large and small $\Wi$. For small $\Wi$, the max trace stress scales
linearly with $\Wi$:
\begin{equation}
  \max \tr\sig \sim \left(\frac{2\De\sinh\left(\frac{1}{2\De}\right)-1}{\De\sinh\left(\frac{1}{2\De}\right)}\right)\Wi, \quad
  \textrm{for } \Wi<<1.
\end{equation}
For large $\Wi$, the max trace of the stress to leading order is 
\begin{equation}
  \max(\tr\sig)\sim \left(\frac{1}{\sinh\left(\frac{1}{2\De}\right)}\right) \frac{\exp\left(\frac{\Wi}{2\De}\right)}{\Wi^2},
  \quad \textrm{for } \Wi>>1.
\end{equation}
This expansion is generated by assuming not only that
$\Wi$ is large, but also that $\Wi$ is large compared to $\De$.
Thus the transition from the low Weissenberg number regime to the high
Weissenberg number regime depends on the Deborah number.  In
Figure \ref{fig:LAOE_sqwv}(a) we include plots of the asymptotic
expressions for the stress for both high and low $\Wi$ for
$\De=0.5$. The transition for the linear behavior to the exponential
behavior appears to occur somewhere between $\Wi=2$ and $\Wi=4.$

For fixed $\Wi$, the stress is a decreasing function of $\De$. 
\change{The case of $\De=0$ corresponds to the steady extension
case. 
 For
high $\De$, the duration of stretch is short, and the stresses do not
get very large before the flow changes to compression.}  

With this nondimensionalization both $\Wi$ and $\De$ scale with the
relaxation time. In studies of locomotion, it is useful consider how
the speed depends on $\De$ for a given gait.  Similarly, in Section
\ref{flexor:sec}, we consider how the stress depends on relaxation
time as an object changes shape with a fixed amplitude. For such
problems, it is useful to consider how the stress depends on the
nondimensional stretch rate $\mathcal{A}=\alpha T=\Wi/(2\De)$
%
 in place of $\Wi$. This other
parameter captures the amplitude of the stretching independent of the
relaxation time, and the relaxation time only appears in
$\De$. For this nondimensionalization $\De=0$ corresponds to Newtonian
flow.

In Fig. \ref{fig:LAOE_sqwv}(b) we plot the max of the trace of the
stress a function of $\De$ for a range of stretch rates $\mathcal{A}
=1,2,4$.  As $\De$ goes to either $0$ or $\infty$, the stress goes
to zero, and as a result the stress is a nonmonotonic function of
$\De$. The $\De$ where the peak stress occurs is fairly insensitive to
the stretch rate \change{within the ranges of $\mathcal{A}$ presented.}

\change{In \ref{C} we present results of a numerical solution to Eqs. \eqref{sig11ode}--\eqref{sig22ode}
for the Giesekus model to demonstrate how additional nonlinear terms effect the theory developed in this section. }


\section{Stress response to oscillating extension with coupling}\label{sec_4roll}
The analysis from the previous section considered the flow fixed
independent of the stress. Here we examine an oscillating extensional
flow in which the stress and velocity are coupled. We drive the system
by prescribing a 4-roll mill type background body force and solve for the resulting
velocity and stress numerically and compare the results with those
from the previous section.

We adapt the model problem from
\cite{thomases2007emergence,guy2015computational} in which the stress
at steady extensional points was examined numerically.  Specifically,
we solve the Stokes-Oldroyd-B 
equations, Eqs. \eqref{stokes}-\eqref{ucderiv} on the two-dimensional periodic domain $[-\pi,\pi]^{2}$ with
a driving background body force
\begin{equation}\label{4roll_F}
  \f = 2\alpha\sin\left(2\pi t/T\right)
    \left(\begin{array}{l}
                -\sin x \cos y \\
      \phantom{-}\cos x \sin y
   \end{array}\right).
\end{equation}
Note in the stokes limit ($\lambda=0$) this body force drives the flow
$\u=-\f/2$. At the origin the linearized flow is identical to the flow
defined in equation \eqref{oscext1} from the decoupled problem with $h(t) =
\sin\left(2\pi t \right)$.

The system is solved with a pseudo-spectral method for spatial
derivatives and a 2nd order implicit-explicit time integrator; small
stress diffusion is added to control stress gradients. See \ref{A} for
details on the numerical method and discretization parameters.

%

As before, the Deborah number is $\De=\lambda T$. The Weissenberg
number is computed as $\Wi=\lambda\gdot$ where
$\gdot=\max_{t}\sqrt{2\srt:2\srt/2}$ for $\srt$ measured at the origin
once the solution has equilibrated to the periodic solution.  For a
Newtonian fluid $\gdot=2\alpha,$ but we find that even for
viscoelastic fluids in the highly nonlinear regime of large stresses
$\gdot\approx2\alpha$. Hence for this problem one can use $\Wi =
2\alpha\lambda$ and nondimensionalize the polymer stress by
$\sig=\tauB/(2\mu_p\alpha)$,
which is equivalent to what was done in the previous section.

\begin{figure}
  \centering
  \includegraphics[width=0.95\textwidth]{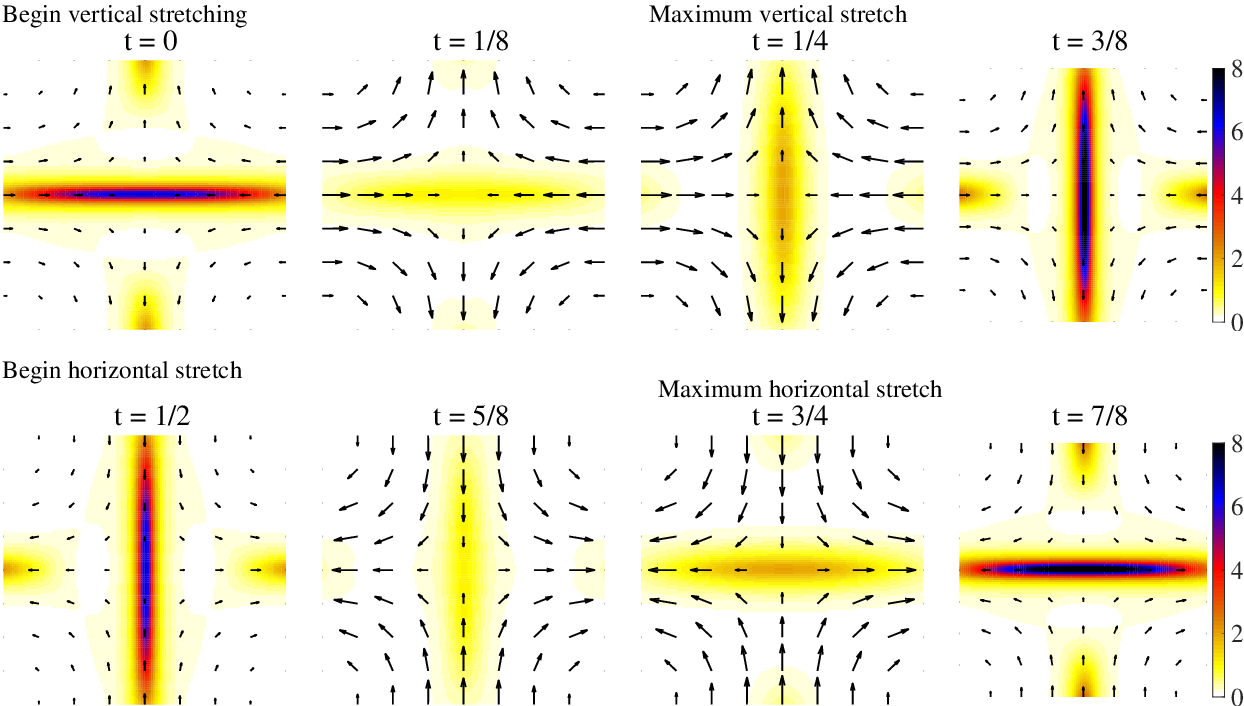}
  \caption{Velocity field overlayed on colorfield displaying polymer
    strain energy density over one period for oscillating extension
    near the origin on the domain $[-\pi/2,\pi/2]^2$ for the fully
    coupled 4-roll mill simulations, at $\De=0.5,$ $\Wi = 12.$}
  \label{fig:4roll_OE}
\end{figure}

In Fig.\ \ref{fig:4roll_OE} we show $\tr\sig$ near the origin (on part of the simulation domain: $[-\pi/2,\pi/2]^2$) for
8 different times over one period for $\Wi=12$ and $\De = 1/2$ after
the solution has reached the periodic solution. These plots show the large
stresses switching orientation over the course of a period due to the
alternating directions of stretching. At the beginning of the period
(t=0) the stress is oriented in the horizontal direction, but at this
time, the flow begins stretching in the vertical direction and
compressing in the horizontal direction. Over the next half period,
the stress becomes increasingly oriented in the vertical
direction. Over the second half of the period, the flow stretches in
the horizontal direction and compresses in the vertical direction.

\begin{figure}
  \centering
  \includegraphics[width=0.95\textwidth]{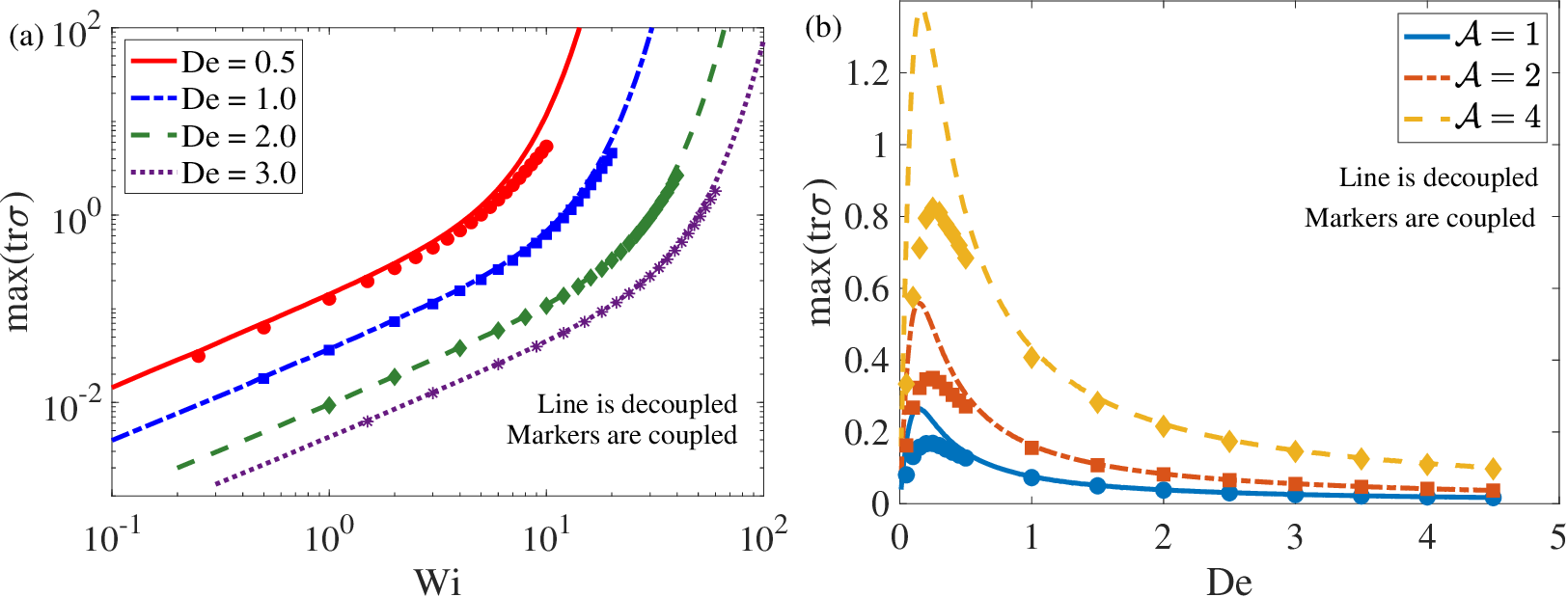}
  \caption{ Max polymer strain energy density at the origin in the
 decoupled ODE (line) and the 4-roll mill coupled simulations (markers) as a function of $\Wi$ for fixed $\De$  (a), and as a function of $\De$ for fixed $\mathcal{A}$ in
    (b).}
  \label{fig:LAOE_4roll}
\end{figure}

To compare the results of the fully coupled simulations with the theory presented in Sec.\ \ref{sec_theory} we 
 numerically solve the ODE's in Eqs.\ \eqref{sig11ode}-\eqref{sig22ode} with temporal oscillation 
$h(t)=\sin(2\pi t).$  \change{We note that there is no closed form analytic expression such as Eq. \eqref{sig_theory} for 
a sinusoidal oscillation.} This is the  analog of the coupled 4-roll mill at the origin, but in the decoupled limit,
i.e.\ where the stress does not affect the velocity.
In Fig.\ \ref{fig:LAOE_4roll} we examine plots of $\max(\tr\sig)$
 as a function of dimensionless parameters in both the coupled case as well as in the decoupled case. 
 The behavior of the stress with sinusoidal temporal oscillation
 is qualitatively the same as that of the square-wave oscillation. As
before, the stress as a function of $\Wi$ shows two regimes: a low
$\Wi$ regime in which the stress is a linear function of $\Wi$, and a
high $\Wi$ regime in which the stress dependence on $\Wi$ is
exponential. The $\Wi$ at the transition between these two regime
again depends on $\De$. The agreement between the decoupled and coupled cases is 
quite good, \change{but in the high $\Wi,$ low $\De$ regime, where the largest stresses are expected,}
there is disagreement. In this problem we find that the non-linear coupling has the effect of 
reducing the stress rather than enhancing it. This is consistent with what was found in \cite{thomases2007emergence}
where the nonlinearities modified the flow and also reduce the stress.  



\section{Flows around flexing objects}
\label{flexor:sec}
\subsection{Bending, no translation}\label{flex_flex}

\begin{figure}
  \centering
  \includegraphics[width=\textwidth]{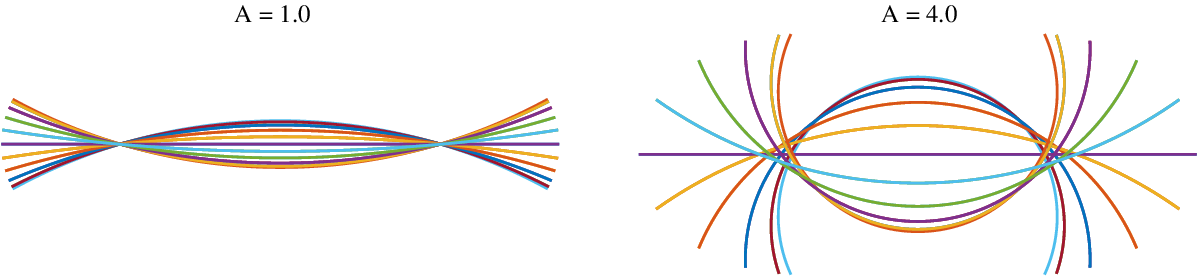}
  \caption{Prescribed shapes for low and high amplitude flexors.}
  \label{fig:fl_shapes}
\end{figure}

In order to simplify the study of the flows around the tips of undulatory swimmers we
consider filaments of length $L=1$ oscillating through circular arcs
with peak curvature $A$. Specifically, the curvature is
\begin{equation}\label{flex_curv}
  \kappa(s,t)=A\sin\left(\frac{2\pi}{T} t\right).
\end{equation}
For $A=\pi$ the fully bent shape is a semi-circle. We consider low
amplitude $A\approx 1$ and high amplitude $A\approx 4$; see
Fig.\ \ref{fig:fl_shapes}.  By symmetry, this motion does not result
in any horizontal translation of the body.
%
%
We refer to these non-translating ``swimmers" as \textit{flexors}. We
previously used these objects to study the effect of viscoelasticity
on soft swimmers in \cite{thomases2017role}.
In what follows we solve Eqs. \eqref{stokes}-\eqref{ucderiv}, and the
external force density, $\f$ is used to enforce the prescribed shape
of the swimmer.
The method is similar to \cite{li2017flagellar} where the shape
is given and the system is solved under the constraint that it is
force and torque free. Thus the flexor has a fixed shape, but it is
free to move in the fluid. Details of the numerical method are given in
\ref{B}.
This method is different from previous swimmer simulations 
\cite{teran2010viscoelastic,thomases2014mechanisms,salazar2016numerical,thomases2017role} which enforced a
prescribed shape approximately using forces that penalized deviations
from a target curvature.

In the flexor model the Deborah number is defined as $\De=\lambda/T,$
where $T$ is the period of motion of the flexor. Defining a
Weissenberg number is more complicated than in the 4-roll
mill. Because the strain rate varies significantly at different places
in the flow, it is not clear how to define a characteristic $\gdot.$
\change{In a region near the tip we find that $\gdot\propto A$ with
constant of proportionality $\approx 2,$ and hence we define $\Wi=2\lambda A,$
and scale the polymer stress as $\sig=\tauB/(2\mu_pA)$.  This is consistent with 
the scalings in Sec. \ref{sec_theory} and \ref{sec_4roll}.  We give
more evidence for $\gdot\approx 2A$ in what follows. }


\begin{figure}
  \centering
  \includegraphics[width=0.95\textwidth]{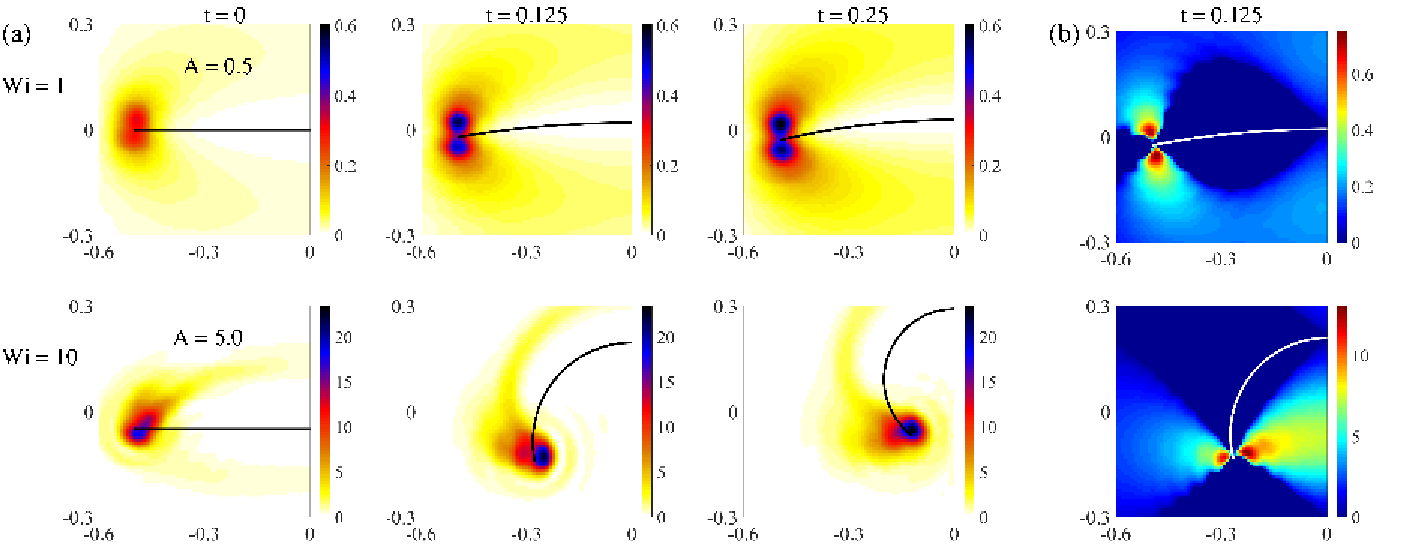}
  \caption{(a) Polymer strain energy density ($\tr\sig$) around flexors at $\Wi=1$ and $\Wi=10$ with $\De=1,$ corresponding to low ($A=0.5$) and high ($A=5.0$) amplitude flexors at different times during a period. (b) Maximum stretch rate $\nu$ for low and high amplitude flexors.}
  \label{fig:fl_sig}
\end{figure}

In Fig.\ \ref{fig:fl_sig} (a) we plot the strain energy density $\tr\sig$
for low amplitude ($A=0.5$) and high amplitude ($A=5)$ flexors during
the downstroke of the motion for $\De=1.$ All results are shown after
the flow has equilibrated to a periodic state at $t\approx 10\lambda.$
We note that the stress is localized at the tips of the flexors during
the motion. It has a much larger scale for the large amplitude
case. The spatial distribution of stress is more symmetric about
the flexor for the low amplitude case than in the high amplitude case.
The low amplitude case corresponds to $\Wi=1$ and the high amplitude
case corresponds to $\Wi=10.$ \change{Suggested by the theory developed for pure oscillating extension, shown in}
Fig.\ \ref{fig:LAOE_4roll}(a), the values $\Wi=1,\;\De=1$ are well in the linear
regime, whereas $\Wi=10,\De=1$ are in the transition region between linear
to exponential.

To understand why the stress concentrates preferentially near the tips
of the flexors we examine the flow near the tips.
In \cite{li2019orientation} we showed that stretching near the tips of
translating cylinders led to large concentrated stresses beyond a
critical $\Wi$. To identify the critical $\Wi$, we identified
a quantity called the maximum stretch rate as the maximum real part of the
eigenvalues of the operator $\Str[\u]\C\equiv\left[\nabla\u\C+\C\nabla\u^T\right].$
The term $\Str[\u]\tauB$ arises in the upper-convected Maxwell
equation, Eq.\ \eqref{ucderiv}, and the eigenvalues of $\Str[\u]$
define the define the growth (or decay) rates of stress due to
stretching (or compression) along particle paths.
The solution to the eigenvalue problem
$\Str[\u]\C=\nu\C$ is $\C=\v_i\v_j^T,$ $\nu_{ij}=\mu_i+\mu_j,$ where
$\mu_i$ is an eigenvalue of $\nabla\u$ with corresponding eigenvector
$\v_i.$ We define the maximum stretch rate $\nu$ at a point defined by
\begin{equation}\label{stretchrate}
  \nu = 2\max(\textrm{Re}(\Lambda(\nabla\u))),
\end{equation}
where $\Lambda(A)$ is the set of eigenvalues of the matrix $A.$
The max stretch rate is related to the shear rate $\gdot$, but it
quantifies specifically the rate of local extension where the
nonlinearities in the stress evolution equation are significant.
In Fig.\ \ref{fig:fl_sig} (b) we plot the max stretch rate for low and
high amplitude motion when the flexor is in the middle of the
downstroke. It is clear that the highly extensional regions of the
flow are at the tips for low and high amplitude flexors.

\begin{figure}
  \centering
  \includegraphics[width=0.85\textwidth]{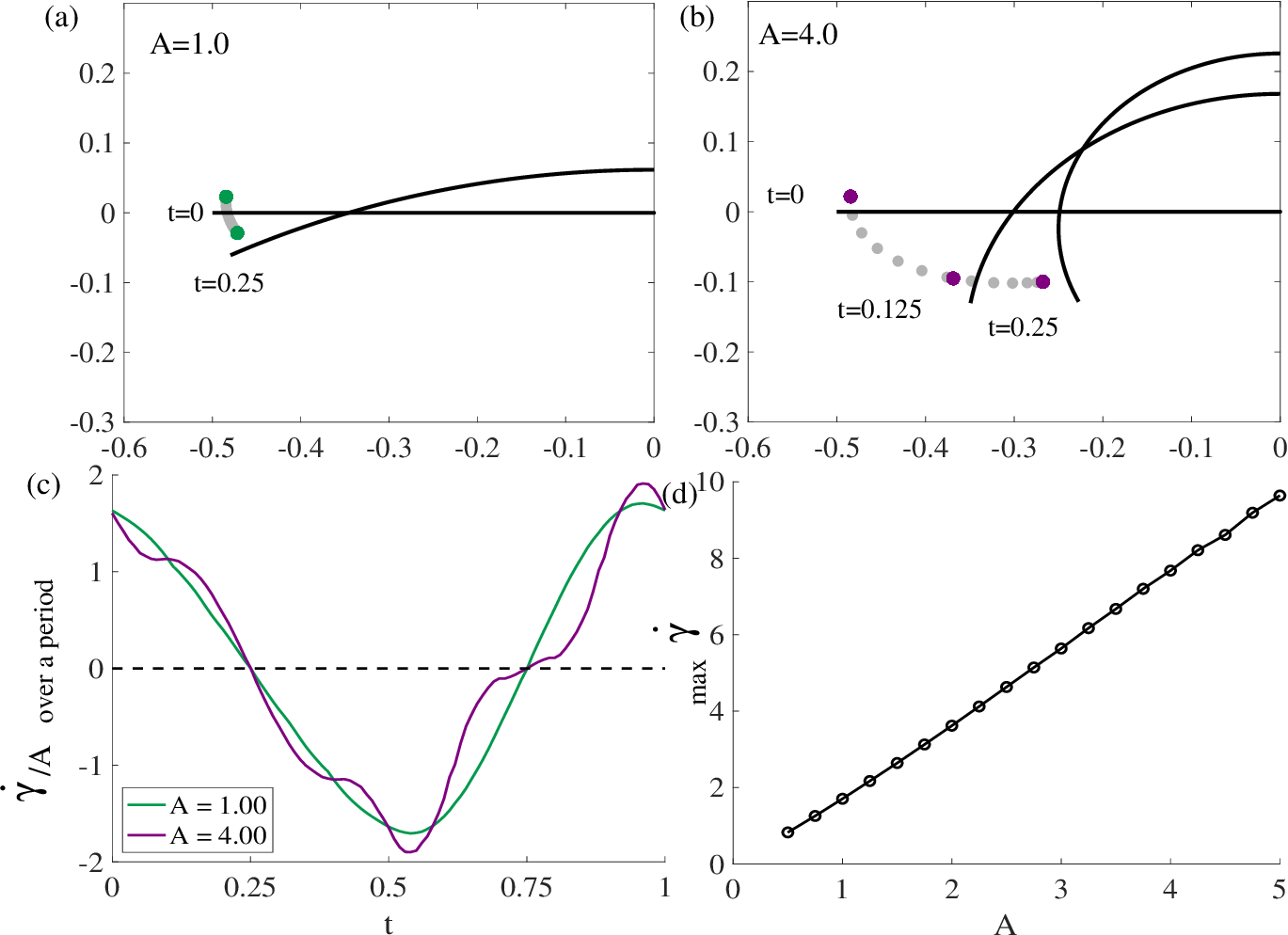}
  \caption{Flexor motion and particle path for point near tip for
    $A=1$ (a) and $A=4$ (b) flexors in a Newtonian fluid. (c) Rate of strain (averaged near
    tip of flexor) over a period scaled by $A$ for $A=1,4$ in a Newtonian fluid. (d)
    Maximum shear rate averaged over a region near the tip is
    proportional to the amplitude with proportionally constant
    approximately 2.
   }
  \label{fig:maxSR_flex}
\end{figure}

We examine the flow near the tips by following particle paths in the
Newtonian flow. In Fig.\ \ref{fig:maxSR_flex} (a)-(b) we highlight a
portion of the particle path near the tip for low and high amplitude
flexors.  To measure the strain rate near the tip we first take the
average $\gdot$ over a set of trajectories which begin at $t=0$ in a
square just above the tip of the flexor of width $\approx .05L.$ In
Fig.\ \ref{fig:maxSR_flex} (c) we plot this average $\gdot(t)/A$ over
one period.  In this region the strain rate is oscillating
periodically between $\pm 2A,$ i.e. $\gdot\sim 2A\sin(t/T).$ The
temporal maximum of $\gdot$ averaged in the neighborhood near the tip
is plotted in Fig.\ \ref{fig:maxSR_flex} (d) over a range of
amplitudes.  A linear fit to the data gives a slope of $1.98.$

\begin{figure}
  \centering
  \includegraphics[width=\textwidth]{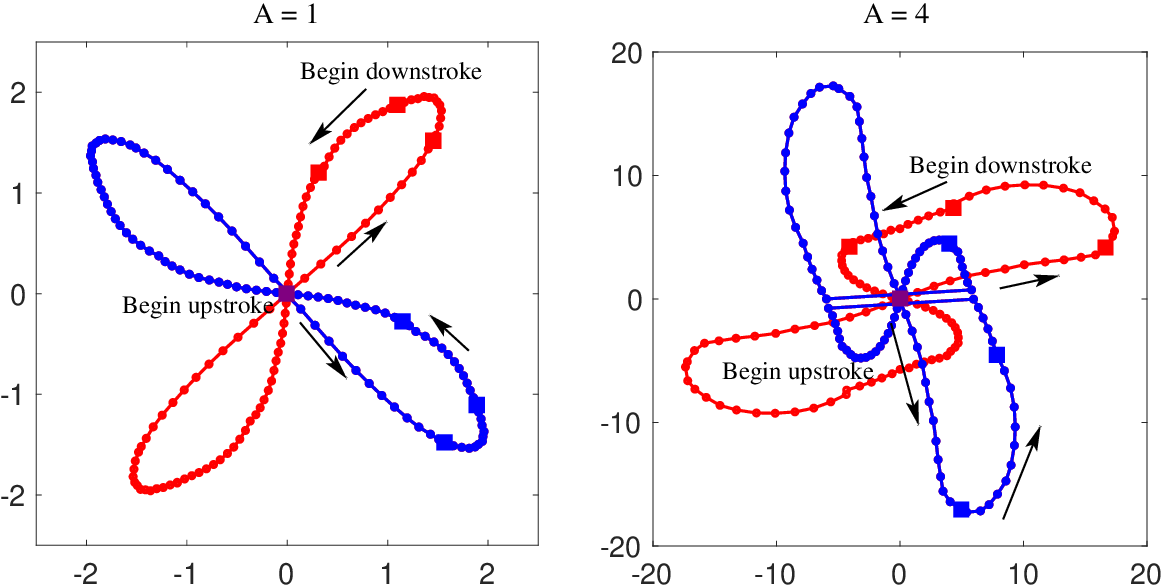}
  \caption{Amplitude and direction of principle stretch given by
    eigenvectors and eigenvalues of $\srt$ in a Newtonian fluid over a period along a
    particle path near the tip for low and high amplitude flexors.}
  \label{fig:fl_flower}
\end{figure}

To examine the direction of stretching, in Fig.\ \ref{fig:fl_flower}
we plot the eigenvectors with positive eigenvalue of the strain rate
tensor, $\srt$, scaled by the eigenvalue over a period on a particle
trajectory that begins slightly above the flexor.  In these plots the
dots are equally spaced in time and hence indicate speed of
motion. The red dots correspond to the down stroke and the blue are
the upstroke. The three additional highlighted times in each portion
of the motion correspond to eighths of a period, with $t=1/4, 3/4$
when the flow is at rest.  It is notable that the direction of
stretching on the downstroke is perpendicular to the direction of
motion on the upstroke, which is indicative of an oscillating
extensional flow. Unlike the problems analyzed in the previous
sections, there is some rotation of the stretching direction. Also
note that the high amplitude case has a more complicated path as well
as a longer path relative to the amplitude. Although the trajectories
change for different initial position these are representative of a
region of points in the fluid above (or below) the flexor near the
tip. Thus we conclude that near the tip the fluid particles are
experiencing an oscillating extension in this region along with some
rotation.
%

  To compare the results of
stress response for flexors with the analytic solution and 4-roll
simulations we quantify the polymer stress around the flexor tip.  We define $\max(\tr\sig)$
by averaging over trajectories, similar
to how we defined  $\gdot$ in Fig. \ref{fig:maxSR_flex} (d). For $\sig$ we choose
the location of the patch of fluid over which the average is taken to be centered on
 the spatio-temporal maximum of $\sig,$ and we use a patch size $0.1L.$ \change{(Note that now we average over a patch 
 size that is double the size used to define the stretch rate  to account for the symmetry breaking that occurs with fluid elasticity. We include 
 regions above and below the tip.) } With the
 region specified, we define $\max(\tr\sig)$ as the spatial average of the maximum
in time of $\tr\sig$ on this region.

\begin{figure}
  \centering
  \includegraphics[width=0.95\textwidth]{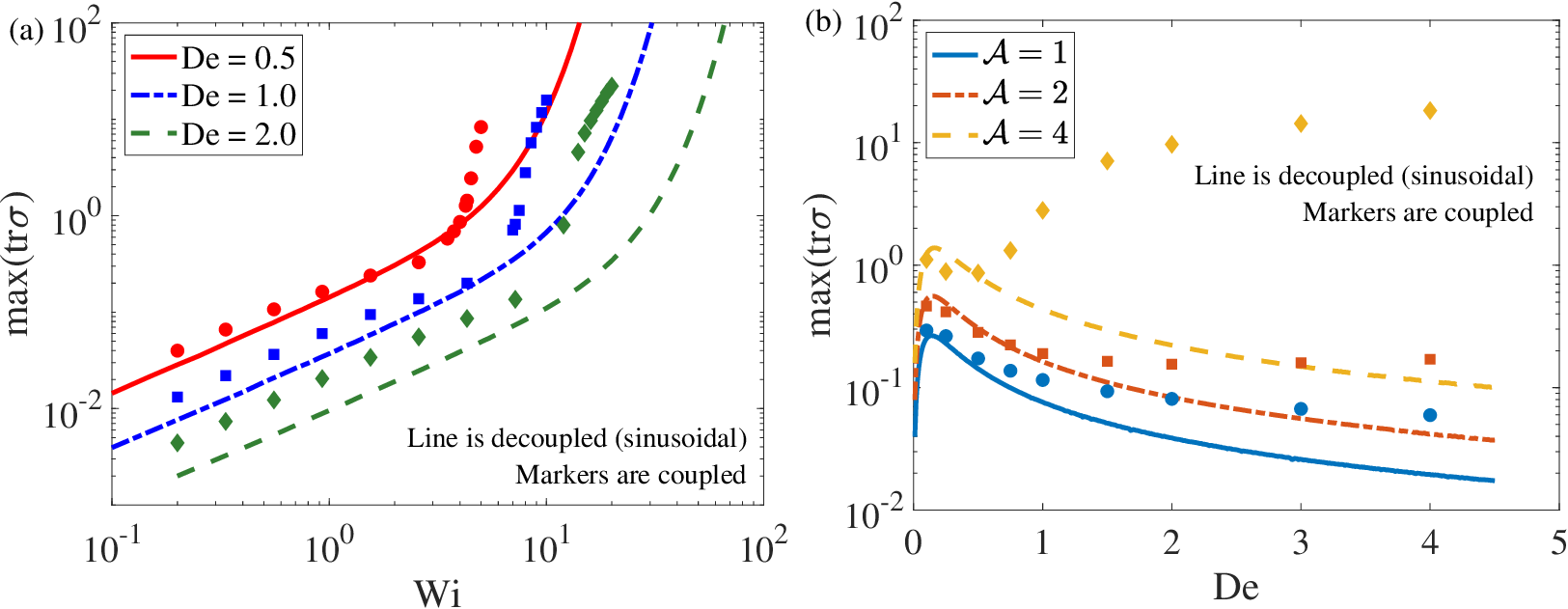}
  \caption{Polymer strain energy density for flexors (markers) and
    decoupled theory with sinusoidal forcing (lines). (a) Stress
    dependence on $\Wi $ for a range of $\De$ (b) Stress dependence on
    $\De$ for a range of $\mathcal{A},$ the non-dimensional ``gait".}
  \label{fig:LAOE_fl}
\end{figure}

We plot $\max(\tr\sig)$ for a range of $\De$ and $\Wi$ in
Fig.\ \ref{fig:LAOE_fl}. In order to compare with the theory from
Sec.\ \ref{sec_4roll} we also plot the theoretical predictions for the
stress at the oscillating extensional point on
Fig. \ref{fig:LAOE_fl}. The solid lines come from the theory for the
decoupled oscillating extensional flow with sinusoidal temporal
forcing, and the markers are results of simulations of flexors.

%

The stress as a function of $\Wi$ from Fig.\ \ref{fig:LAOE_fl} 
again shows linear behavior at low $\Wi$ and exponential at high $\Wi$
for the flexor. At low $\Wi$ and low $\De$ the stress is similar to
the decoupled theory, but generally the stress response is significantly larger
for the flexor simulations than from the theory.  Note that while the
stress response is stronger for the flexors than the theory
predicts, the theory is still able to capture the location of the
transition fairly well; the large stress growth appears near the bend
in the theory curve.

The deviation from the decoupled theory (and 4-roll mill simulations)
is more pronounced when examining the stress as a function of $\De,$
for a range of $\mathcal{A}=\Wi/(2\De)$, which, as before, is
the nondimensional stretch rate and is proportional to the amplitude.
As shown in Fig.\ \ref{fig:LAOE_fl} (b), for low amplitude there is
qualitatively similar stress dependence on $\De$ for different
$\mathcal{A}$. However, for the high $\De$ regime the stress is much
larger than in the theory or in the four-roll simulations. Notably,
for the highest amplitude, the stress is increasing as a function of
$\De$, which is a fundamentally different behavior than in the other
problems.

\begin{figure}
  \centering
  \includegraphics[width=0.95\textwidth]{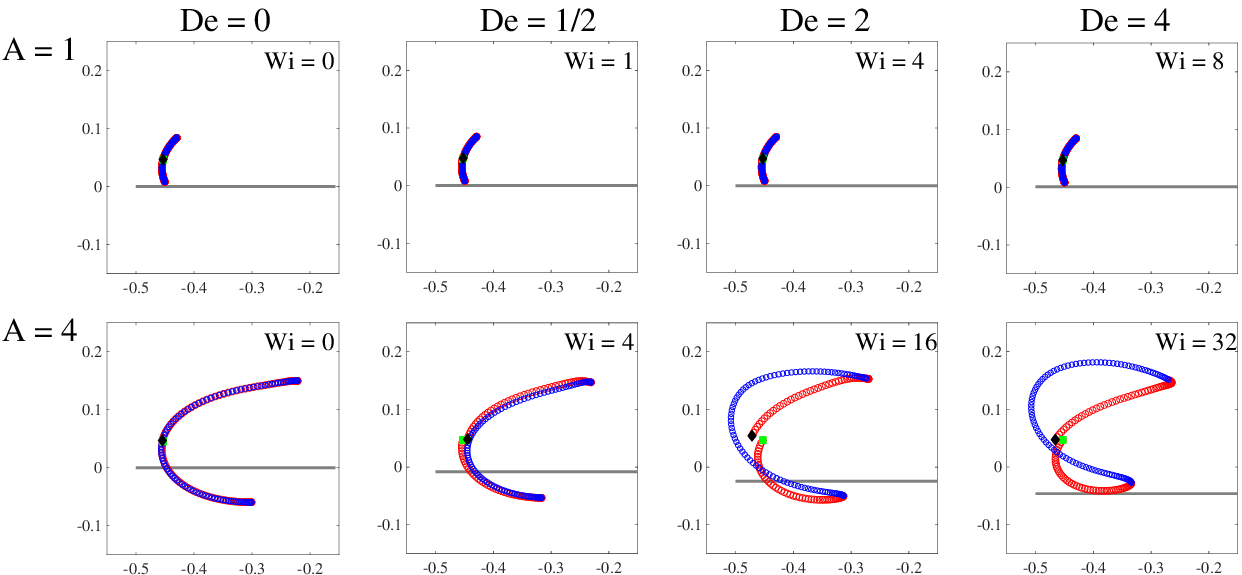}
  \caption{Particle trajectories over one period for flexors at low and high amplitude. Particles start at green diamond and end at black square. Color indicates when the flexor is in the upstroke (red) or downstroke (blue). Grey line is flexor initial position. }
  \label{fig:loops}
\end{figure}

To demonstrate why the high amplitude/high $\De$ flexors exhibit such
different stress response than the theory, in Fig.\ \ref{fig:loops} we
plot the trajectories of points near the tips of the flexors for low
and high amplitudes at a range of $\De$.  The $\Wi $ range here is
$\Wi=0,1,4,8$ for low amplitude and $\Wi =0, 4,16, 32$ for high
amplitude.  In the large amplitude case, the feedback on the flow from
the viscoelastic stresses changes particles paths to make non
time-reversible trajectories. The results is that the fluid particles
do not feel equal stretching/compression as they do when the path is
time-reversible. These fluid particles no longer experience mean zero
stretching, and the result is large stress accumulation. Thus the high
$\De,$ high $\Wi$ deviation from the theory is a result of nonlinear
feedback, which is very different from the affect of such feedback in
the four-roll mill simulations where the extensional point was fixed
in space.

\subsection{Undulatory Swimmer}\label{flex_swim}

\begin{figure}
  \centering
  \includegraphics[width=0.95\textwidth]{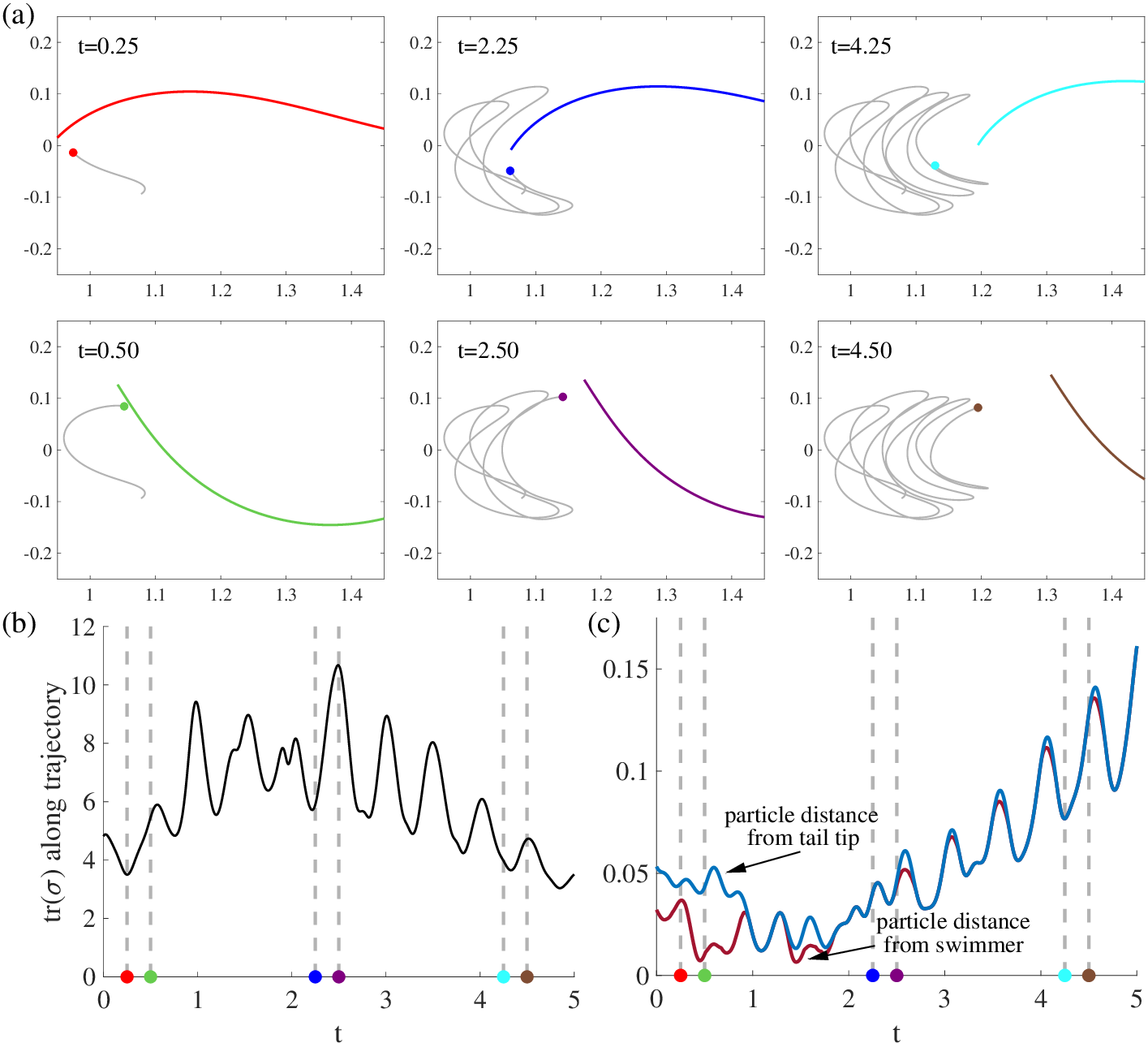}
  \caption{(a) Large amplitude swimmer ($(A_t,A_h)=(5,2)$) at 6 snapshots during 5 periods of motion (swimmer is moving left to right, frame is held fixed). Particle path-lines indicate position of particle in the flow over time. (b) Trace of polymer stress at the particle point indicated in (a) over 5 periods. (Snapshot times indicated by corresponding color and dashed lines.)  (c) Distance from particle to tail tip and distance from particle to swimmer over time. }
  \label{fig:SW}
\end{figure}

Here we explore the relevance of the previous results to a swimmer which is bending and translating.
 It has been observed that large polymer stress islands concentrate near tips of swimmers \cite{teran2010viscoelastic,thomases2014mechanisms,thomases2017flexible}.
We demonstrated this stress concentration in Fig. \ref{fig:TWALR}(b), and using the 
observations from Sec. \ref{flex_flex} that relate curvature to $\Wi,$ we compute $\Wi=2\mathcal{A}\De\approx 5,$ and $\Wi\approx 20$ for $\De = 0.5$, and $2$ respectively, based on tail curvature amplitude $A_t=5.$  
%
%
%
%
In Fig. \ref{fig:SW} (a) we show 6 snapshots of the large amplitude swimmer with a representative particle and path-line over 5 periods as the particle moves around the tip. This swimmer has the same gait as the swimmer in Fig. \ref{fig:TWALR}(b), and here $\De=2.$
%
%
In Figs. \ref{fig:SW} (b)-(c) we plot the polymer stress at that particle point (in (b)) and both the distance between the particle and the tip as well as the distance between the particle and the swimmer (in (c)) over the 5 periods. 
We note that over these 5 periods the polymer stress increases as the particle moves closer to the tip and decreases as the particle moves away from the tip. 
%
%
%
When translation is included any single particle only remains near the tip for a few periods, however there are new particles entering the tip region and experiencing the large stretch and this is sufficient to maintain a large polymer stress patch in the tail region over  time.

\section{Conclusions}

We extend the well-known $\Wi$ transitions in steady extension to
oscillatory extension, and unlike steady extension we find that
bounded solutions exist for all $\Wi,$ but there is a $\De$ dependent
$\Wi$ transition beyond which the size of the stress grows
exponentially in $\Wi.$
In simulations of swimmers in the high amplitude, high $\De$ case from
Fig.\ \ref{fig:TWALR}(b) the swimmer is well in the nonlinear regime
at $\Wi=20$ and $\De=2.$
Comparing the stress as a function of $\De$ for swimmers and flexors
from Fig.\ \ref{fig:TWALR}(c) and Fig. \ref{fig:LAOE_fl}(b),
respectively, shows a similar amplitude dependent response, which is
different from the theory and simulations of stationary extensional
points. 
%
%
Previous simulations \cite{teran2010viscoelastic,thomases2017flexible}
noted that swimming speed dependence on $\De$ was different for high
and low amplitude gaits. Here we explain these observations by
identifying a $\Wi$ transition which shows that high amplitude gaits
operate in the regime of large stress growth.

Both the oscillatory extension stagnation point theory and the 4-roll
mill simulations exhibited non-monotonicity in the stress response for
a fixed ``gait" ($\mathcal{A}=\Wi/(2\De)$). This non-monotonicity
comes from the fact that in the limit as $\De\rightarrow 0,$ the
stress must go to zero, and as $\De\rightarrow\infty$ oscillations are
averaged out and the stress again goes to zero. For flexors we saw a
similar non-monotonicity at low amplitude, but at higher amplitudes
we saw that particle paths were deformed and thus the stress growth
and decay were no longer averaging out.   \change{This
non-monotonicity appears to be related to non-monotonic speed
responses that have been observed, but not explained, in simulations  
\cite{teran2010viscoelastic,thomases2014mechanisms,salazar2016numerical,thomases2017flexible}
(also shown in Fig. \ref{fig:TWALR}(d)).
}  



For the flexors at high $\De,$ the stress growth and decay do not
average out as they do in the four-roll mill because the systems are
driven differently.  In the 4-roll mill system, the steady extensional
point is driven by a background force, and as the stress grows the
velocity nearby changes in such a way that reduces the stress.  By
contrast, for the flexors, the motion is fixed, but the location of the
extension is free to move, and as the extension moves fluid patches no
longer feel an equal stretch-compress. In the 4-roll mill the
nonlinear feedback actually weakens the stress where with the flexors
the nonlinearities are amplified.
  
These results apply to planar motion, but 3D simulations of slender
objects in viscoelastic fluids have also demonstrated large stress
concentrating near tips
\cite{li2017flagellar,li2019orientation,guido2019three}. In \cite{guido2019three}
undulatory swimmers have been simulated in a 3D viscoelastic fluid and
it was found that the swimming speed does not decay as rapidly with
$\De$ as was seen in 2D \cite{thomases2014mechanisms}. We believe
there will still be a $\Wi$ transition for undulatory motion in 3D,
but the quantitative results on stress accumulation and the
implications on swimming may depend on the spatial dimension.

\section*{Acknowledgement}The authors thank David Stein for helpful discussions on this work. R.D.G. and B.T. were supported in part by NSF Grant No.\ DMS-1664679. 

\appendix
\section{Numerical Method: 4-roll mill} \label{A}

For the 4-roll mill simulations we solve
Eqs.\ \eqref{stokes}-\eqref{ucderiv}, with forcing given by
Eq.\ \eqref{4roll_F}.  The fluid domain is a 2D periodic box of length
$2\pi.$ We use $\Delta x=2\pi/256\approx 0.024,$ for the fluid
discretization, and fix the viscosity ratio $\mu_p/\mu_s=0.5.$ We
use a pseudo-spectral method for spatial derivatives and evolve the
conformation tensor $\C,$ which is related to the polymer stress
tensor through $\tauB=\mu_p/\lambda(\C-\Id).$ The conformation
tensor evolves according to
\begin{equation}
  \label{ucderiv_alt}
  \C+\lambda\stackrel{\nabla}\C =\Id+\eta\Delta\C,
\end{equation}
where polymer stress diffusion is added as numerical smoothing
\cite{sureshkumar1995effect,thomases2011analysis}.  The diffusion
coeffient used is $\eta=c\Delta x^2,$ so that as $\Delta x\rightarrow
0$ the model converges to the Oldroyd-B model.  In these simulations
$c=2$ and the artificial diffusion does not effect the qualitative
results reported here.

We use the Crank-Nicholson-Adams-Bashforth second order
implicit-explicit time integrator to evolve the conformation tensor,
$\C.$ The time-step we choose depends on the amplitude $\alpha$ but
ranges between $\Delta t=0.001$ and $\Delta t =0.0001$ chosen to
maintain stability.

\section{Numerical Method: Flexors} \label{B}
For the flexor simulations we solve the fluid-structure equations
Eqs.\ \eqref{stokes}-\eqref{ucderiv}, where the forcing term $\f$
results from the prescribed motion of the flexor.  We use a method
similar to that from \cite{li2017numerical}. The shape, and hence
velocity, of the flexor is prescribed in a fixed body frame.  The
position of the flexor in the lab frame is given by $\X(s,t) =
\X_{p}(s,t) + \X_{0}$, where $s$ is a Lagrangian on the body,
$\X_{p}(s,t)$ is the prescribed shape in fixed a body fixed frame, and
$\X_{0}$ is the translation of the origin in the body frame to the lab
frame. The velocity of the body is $\U = \U_{p}+\U_{0}$, where
$\U_{p}=\partial_{t}\X_{p}(s,t)$ is the prescribed velocity in the
body frame, and $\partial_{t}\X_{0}=\U_{0}$ is the unknown
translational velocity.

The forces and translational velocity are determined
implicitly by the constraints of the prescribed shape and no net
force on the body. The immersed boundary method is used to interpolate
the fluid velocity to the swimmer and to transfer forces on the flexor
to the fluid.

In each time step of the simulation we alternately advance the
conformation tensor $\C$ and the fluid/structure system. Given the
current velocity field $\u$ we evolve the conformation tensor
according to Eq.\ \eqref{ucderiv_alt} and thus we have the current
polymer stress $\tauB$. With the given stress and velocity of the
structure we simultaneously solve to the fluid velocity, pressure and
fluid forces on the structure which satisfy
\begin{gather}
  -\nabla p+\Delta\u+ \xi \nabla\cdot\tauB+\mathcal{S}\F=0\label{IBst},\\
  \nabla\cdot\u=0\label{IBdiv}\\
   \mathcal{S}^*\u =\U_{p} +\U_{0} ,\label{IBBC}\\
  \int_{\Gamma} \F ds=0.\label{IBFB}
\end{gather}
The operator $\mathcal{S}$ transfers forces on the flexor to fluid and is defined as
\begin{equation}
  \label{spread}
  \mathcal{S}(\F)=\int_{\Gamma}\F(t,s)\delta_{\Delta x}(\x-\X(t,s))ds,
\end{equation}
where $\delta_{\Delta x}$ is a regularized $\delta$-function. The
discrete $\delta$ is the standard four-point function described in
\cite{peskin2002immersed}. The operator $\mathcal{S}^*$ maps the
velocity field on the Eulerian grid to the flexor body, and is defined
as
\begin{equation}\label{interp}
  \mathcal{S}^*(\u)=\int_{\Omega}\u(t,x)\delta_{\Delta x}(\x-\X(t,s))dx.
\end{equation}

Equation \eqref{IBBC} determines that the structure moves with the
local fluid velocity, i.e.\ there is no slip on the body surface, and
Eq. \eqref{IBFB} enforces the no net force condition on the
structure. These two constraints determine the unknown force, $\F$,
and the unknown translational velocity, $\U_{0}$. To solve this system
of equations we eliminate the fluid velocity and pressure and solve the
smaller system for the body forces and translational velocity
\begin{align}
  \label{SC1}\mathcal{S}^*\mathcal{L}^{-1}\mathcal{S}\F+\U_{0} & = -\U_{p}-\mathcal{S}^*\mathcal{L}^{-1}\nabla\cdot\tauB,\\
  \label{SC2} \int_{\Gamma}\F \, ds & = 0.
\end{align}
Here $\mathcal{L}$ is the Stokes operator that maps a fluid velocity
to the applied forces. After solving for the force on the swimmer we
update the body position in lab frame and the fluid velocity to
complete the time step.


For the flexor simulations our fluid domain is a 2D periodic box of
length $2L,$ where $L=1$ is the flexor size. We use $\Delta
x=1/128\approx 0.008,$ for the fluid discretization and discretize the
flexor with $ds=0.75 dx.$ A Fourier discretization of the spatial
operators is used. Equations \eqref{SC1}-\eqref{SC2} are solved using
the conjugate gradient method, which is preconditioned using the
method of regularized Stokeslets \cite{cortez2001method} to
approximate the operator $\mathcal{S}^*\mathcal{L}^{-1}\mathcal{S}$.

We evolve the conformation tensor using a
Crank-Nicholson-Adams-Bashforth scheme, with a diffusion coefficient
$\eta=9\Delta x^2.$ The time-step we choose depends on the amplitude
flexor but ranges between $\Delta t=0.001$ and $\Delta t =0.0001,$
chosen to maintain stability.

\section{Comparison of ODE model for Oldroyd-B and Giesekus} \label{C}
\newcommand{\galp}{\alpha_g}

 The Giesekus model \cite{giesekus1982simple} is a modification of the Oldroyd-B model in which an anisotropic drag term is introduced
as a quadratic nonlinearity in the polymer stress. This term introduces an additional (small) non-dimensional parameter $ \galp $. Here we consider how solutions to the ODE  in Eqs. \eqref{sig11ode}-\eqref{sig22ode} change with this nonlinear modification. 
 The dimensionless equations are
\begin{align} 
  \label{sig11ode_gie} \De\;\frac{d}{dt}\sig_{11}+(1-h(t)\Wi)\;\sig_{11}+\galp\Wi\;\sig_{11}^2&=\phantom{-}h(t) \\ 
  \label{sig22ode_gie} \De\;\frac{d}{dt}\sig_{22}+(1+h(t)\Wi)\;\sig_{22}+\galp\Wi\;\sig_{22}^2&=-h(t), 
\end{align}
where, as before, the Weissenberg number is $\Wi = 2\alpha\lambda,$ and
the Deborah number is $\De=\lambda/T$. 
Here we use a square-wave profile for $h(t),$ to compare with the results in Fig. \ref{fig:LAOE_sqwv}. 

\begin{figure}
  \centering
  \includegraphics[width=0.65\textwidth]{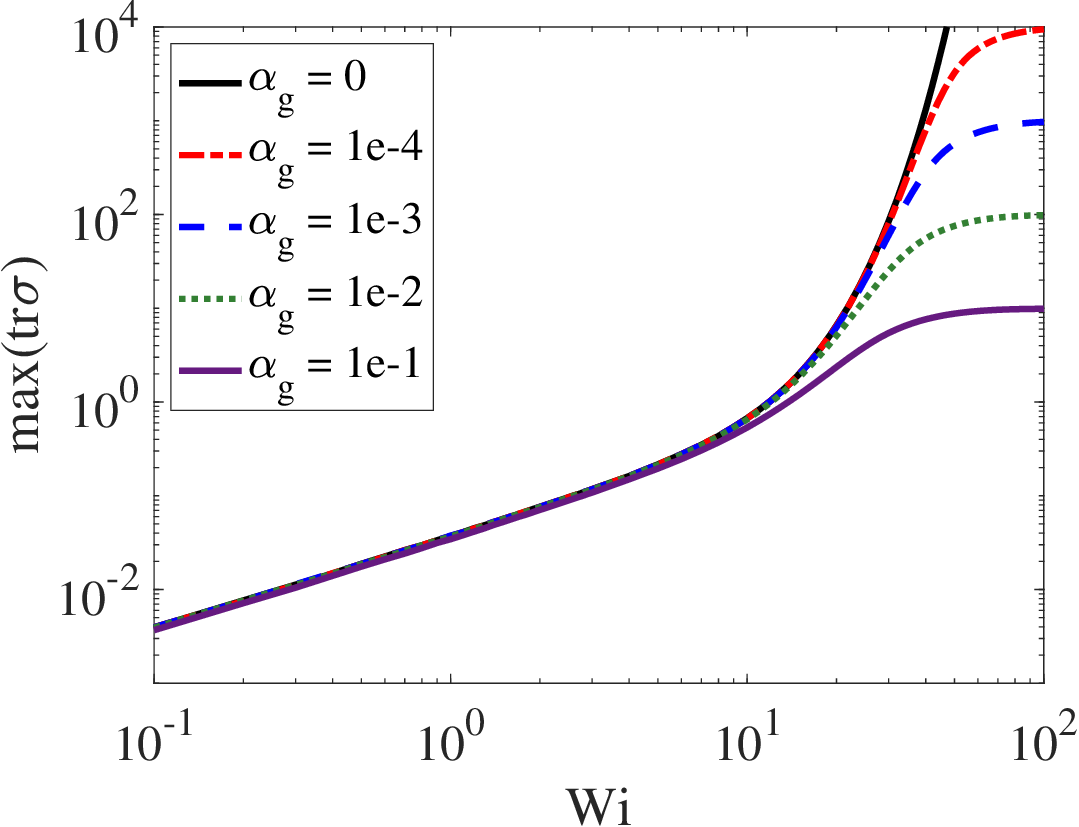}
  \caption{Numerical solution of $\max(\tr\sig)$ as a function of $\Wi$ at $\De = 1,$ for a range of $0\le \galp\le 0.1,$ at a given oscillatory extensional stagnation point in the Giesekus model. }
  \label{fig:gie}
\end{figure}

In Fig. \ref{fig:gie} we plot $\max(\tr\sig)$ as a function of $\Wi$ at $\De=1,$ and compare various values of $\galp,$ including $\galp=0$ (Fig. \ref{fig:LAOE_sqwv}). 
As before, this plot shows two different regimes for
how the stress depends on $\Wi$, and there is a Deborah number dependent
transition between the two regimes (dependence on $\De$ not shown for simplicity). 
The parameter $\alpha_g$ has the effect of bounding $\max(\tr\sig)$ at large $\Wi$, and we note
that the peak stresses are approximately $\galp^{-1}.$  In this problem the value of $\max(\tr\sig)$ in the transition region is in the range of $1-10,$ 
hence for smaller values of $\galp$ the deviations from the Oldroyd-B model
occur after the linear to exponential transition.
%
%
%
%



\end{document}